\documentclass[12pt]{article}
\usepackage{amssymb}
\usepackage{amsmath}
\usepackage{hyperref}
\begin{document}
\title{\bf  Does a Cloud of Strings Affect Shear Viscosity Bound?}

\author{Mehdi Sadeghi$^1$\thanks{Corresponding author: Email: mehdi.sadeghi@abru.ac.ir}  \hspace{2mm} and
      Hadi Ranjbari$^{2}$\thanks{Email: fhranji@gmail.com }\hspace{2mm}\\
		 {\small {\em  $^1$Department of Physics, School of Sciences,}}\\
		 {\small {\em  Ayatollah Boroujerdi University, Boroujerd, Iran}}\\
 {\small {\em   $^2$Department of  Physics, Payame Noor University (PNU),}}\\
        {\small {\em  P.O.Box 19395-3697, Tehran, Iran}}
    }

\date{\today}
\maketitle

\abstract{The Einstein AdS black brane with a cloud of strings background in context of massive gravity is introduced. There is a momentum dissipation on the boundary because of graviton mass on the bulk. The ratio of shear viscosity to entropy density is calculated for this solution. This value violates the KSS bound if we apply the Dirichlet boundary and regularity on the horizon conditions. Our result shows that this value is independent of the cloud of strings.}\\



\section{Introduction} \label{intro}

Graviton is massless in general theory of relativity (GR) formulated by Einstein. However, Hierarchy problem and the brane-world gravity solutions \cite{Dvali2000,Dvali2000-} predict the existence of massive graviton. There are some problems such as the cosmological constant problem and the current acceleration that GR can not explain them. For these reasons GR must be modified. There are some modifications of GR theory such as massive gravity \cite{deRham:2010kj}, bi-metric gravity\cite{Hassan:2011ea}, brane-world cosmology \cite{Gergely2006}, scalar-tensor gravity\cite{Brans1961}, $f(R)$ gravity \cite{Akbar} and Lovelock gravity \cite{Lovelock:1971}.\\ 
Massive gravity formulated in flat spacetime by Pauli and Fierz \cite{Fierz:1939ix} and it's ghost-free and the generalization in the non-flat background was introduced by de Rham, Gabadadze and Tolley (dRGT) \cite{deRham:2010kj} in which the Boulware-Deser  ghost \cite{Boulware72} does not appear.  The accelerated expansion of the universe without considering the dark energy could be explained by massive gravity. Massive gravity is also help us to understand the quantum gravity effects\cite{Vasiliev96}.
In this paper, our goal is further explore of a string cloud in the framework of
massive theories of gravity.\\
Gauge-Gravity duality relates two different theories and it is a useful tool for studying strongly coupled theories\cite{Maldacena97,Aharony99,Casalderrey11,Mateos07}. In this duality, gauge theory lives in the boundary of anti-de Sitter space-time and gravity is inhabited on the bulk. In long wavelength limit, Gauge-Gravity duality leads to fluid-gravity duality\cite{Bhattacharyya07,Bhattacharya2011}. It means fluid mechanics is an effective theory of field theory. The degree of freedoms of theory in boundary reduces to energy density  $\epsilon$, thermodynamics pressure density $p$ and fluid velocity $u^{\mu}$. There is a constrained for fluid velocity as $u^2 = -1$. So the number of fluid variables is five and they can be found by Equation of motion,
\begin{align}
&\nabla_{\mu}T^{\mu \nu}=0, \\
&T^{\mu \nu}=\epsilon u^{\mu} u^{\nu} + p P^{\mu \nu}\nonumber,\\
&P^{\mu \nu}=\eta^{\mu \nu}+u^{\mu}u^{\nu}\nonumber,\\
&\eta^{\mu \nu}=(-1,1,1,1)\nonumber.
\end{align}
 and equation of state $\epsilon=-p+Ts$ where $T$ is temperature and $s$ is entropy density. To illustrate the effects of dissipation we should add extra pieces to the $T^{\mu \nu}$,
\begin{align}
& T^{\mu \nu } =(\epsilon +p)u^{\mu } u^{\nu } +pg^{\mu \nu } -\sigma ^{\mu \nu },\\
&\sigma ^{\mu \nu } = {P^{\mu \alpha } P^{\nu \beta } } [\eta(\nabla _{\alpha } u_{\beta } +\nabla _{\beta } u_{\alpha })+ (\zeta-\frac{2}{3}\eta) g_{\alpha \beta } \nabla .u].
\end{align}
\indent where $\eta$, $\zeta $, $\sigma ^{\mu \nu }$ and $P^{\mu \nu }$ are shear viscosity,bulk viscosity, shear tensor and projection operator, respectively \cite{Kovtun2012}.\\
There are several ways to extract the transport coefficients: Kubo formula, pole method, membrane paradigm and the constitutive relation of energy-momentum stress tensor of the dual fluid and corresponding Navier-Stokes equations. Kubo formula and membrane paradigm are determined on horizon.  In this work we are interested in calculating shear viscosity by Green-Kubo formula \cite{Son2007}. \\
\begin{equation}\label{Kubo}
\eta =\mathop{\lim }\limits_{\omega \to 0} \frac{1}{2\omega } \int dt\,  d\vec{x}\, e^{\imath\omega t} \left\langle [T_{y}^{x} (x),T_{y}^{x} (0)]\right\rangle =-\mathop{\lim }\limits_{\omega \, \to \, 0} \frac{1}{\omega } \Im G_{y\, \, y}^{x\, \, x} (\omega ,\vec{0}).
\end{equation}
where $T{^x\,_y}$ are spatial parts of energy-momentum tensor. The value of $\frac{\eta}{s}$ is one of the most interesting results from fluid-gravity duality which takes the value of $\frac{1}{4\pi}$ for all field theories which are dual to Einstein-Hilbert gravity, known as Kovtun, Son and Starinets (KSS) conjecture \cite{Son2007,Policastro2001,Kovtun2004,Policastro2002}. This lower bound is also satisfied for transverse shear viscosity per entropy density for anisotropic black brane \cite{Sadeghi:2019trg} and five-dimensional $F(R)$-gravity
considered as non-perturbative stringy effective action \cite{Nojiri:2010ny}.
 It is supported both by experimental data and theoretical analysis for quark-gluon-plasma. The ratio of shear viscosity to entropy density is proportional to the inverse squared of the coupling of quantum thermal gauge theory. It means the stronger the coupling, the weaker the shear viscosity per entropy density\cite{Kovtun2003}.\\
In this paper we consider massive gravity with a cloud of strings \cite{Letelier:1979ej,Richarte:2007bx,Yadav:2009zza,Ganguly:2014cqa,Bronnikov:2016dhz,Barbosa:2016lse,Herscovich:2010vr,Ghosh:2014pga,Lee:2014dha,Mazharimousavi:2015sfo,Graca:2016cbd} and introduce the black brane solution. Finally, we study the effect of a cloud of strings on the value of $\frac{\eta}{s}$ and suggest some comments about the field  dual to this gravity model. 
\section{The Einstein AdS Black Brane with a Cloud of String Background in Context of Massive Gravity }
 \label{sec2}
The action is given by,
\begin{equation}\label{action}
I=\frac{1}{2}\int{d^4x\sqrt{-g}\Bigg[R-2\Lambda+m^2\sum_{i=1}^4{c_{i}\mathcal{U}_i(g,f)}\Bigg]}+T_p\int_{\Sigma}\sqrt{-\gamma} d\lambda^0 d\lambda^1,
\end{equation}
where $ R $ is the scalar curvature, $\Lambda=\frac{-3}{2l^2}$ is cosmological constant, $l$  is the radius of AdS spacetime, $f$ is a fixed rank-2 symmetric tensor known as reference metric and $m$ is the mass parameter. $ c_i $'s are constants and $ \mathcal{U}_i $ are symmetric polynomials of the eigenvalues of the $ 4\times4 $ matrix $ \mathcal{K}^{\mu}_{\nu}=\sqrt{g^{\mu \alpha}f_{\alpha \nu}} $ 
 \begin{align}\label{U} 
  & \mathcal{U}_1=[\mathcal{K}]\nonumber\\
  & \mathcal{U}_2=[\mathcal{K}]^2-[\mathcal{K}^2]\nonumber\\
  &\mathcal{U}_3=[\mathcal{K}]^3-3[\mathcal{K}][\mathcal{K}^2]+2[\mathcal{K}^3]\nonumber\\
  & \mathcal{U}_4=[\mathcal{K}]^4-6[\mathcal{K}^2][\mathcal{K}]^2+8[\mathcal{K}^3][\mathcal{K}]+3[\mathcal{K}^2]^2-6[\mathcal{K}^4] 
   \end{align}
The square root in $ \mathcal{K} $ means $ (\sqrt{A})^\mu_\nu(\sqrt{A})^\nu_\lambda=A^\mu_\lambda $ and the rectangular brackets denote traces, where the last part is called the Nambu-Goto action of a string and $({\lambda}^{0},{\lambda}^{1})$  is a parametrization of the worldsheet, $T_p$ is a positive quantity and is related to the tension of the string and $\gamma$ is the determinant of the induced metric \cite{Ghosh:2014pga,Lee:2014dha,Mazharimousavi:2015sfo,Graca:2016cbd}
\begin{equation}\label{gamma}
{\gamma _{ab}} = {g_{\mu \nu }}\frac{{\partial {x^\mu }}}{{\partial {\lambda ^a}}}\frac{{\partial {x^\nu }}}{{\partial {\lambda ^b}}}.
\end{equation}
Nambu-Goto action can be written in terms of ${\Sigma ^{\mu \nu }}$ as follows,
\begin{equation}
{S_{NG}} = T_p\int\limits_\Sigma  {\sqrt { - \frac{1}{2}{\Sigma _{\mu \nu }}{\Sigma ^{\mu \nu }}} } d{\lambda ^0}d{\lambda ^1}
\end{equation}
where
\begin{equation}\label{sigma}
 {\Sigma ^{\mu \nu }} = {\epsilon ^{ab}}\frac{{\partial {x^\mu }}}{{\partial {\lambda ^a}}}\frac{{\partial {x^\nu }}}{{\partial {\lambda ^b}}}
\end{equation}
 is the space-time bi-vector. $\epsilon ^{ab}$ is two-dimensional Levi-Civita tensor,  $\epsilon ^{01}=-\epsilon ^{10}=1$ and $\epsilon ^{00}=\epsilon ^{11}=0$.\\
 $\Sigma ^{\mu \nu}$ must be satisfied in the following conditions to form a surface,
\begin{align}
&\Sigma ^{\mu [\alpha } \Sigma ^{\beta \gamma]}=0\\
&\nabla_{\mu}\Sigma ^{\mu [\alpha } \Sigma ^{\beta \gamma]}=0
\end{align}
Where the square brackets denote antisymmetrization in the enclosed indices. We get an useful identity by definition of $\gamma_{ab}$ Eq.(\ref{gamma}) and $\Sigma^{\mu \nu}$ Eq.(\ref{sigma}),
\begin{equation}\label{Identity3}
\Sigma ^{\mu \sigma } \Sigma_{\sigma \tau}\Sigma^{\tau \nu}=\gamma \Sigma^{\nu \mu}.
\end{equation} 
The energy-momentum tensor for a cloud of strings is calculated by variation of the metric as follows,
\begin{equation}\label{T} 
T^{\mu \nu }= {{2\partial \cal{L}} \over {\partial {g_{\mu \nu }}}}=\rho \frac{{{\Sigma ^{\mu \sigma }}\Sigma _\sigma ^\nu }}{{\sqrt { - \gamma } }}.
\end{equation}
where $\rho$ is the proper density of a string cloud and  conservation of the energy-momentum tensor ${\nabla _\nu }{T^{\mu \nu }} = 0$  results in,
 \begin{equation}\label{TS2}
 \nabla_{\mu}(\rho \Sigma ^{\mu \sigma})\frac{\Sigma_{\sigma}\,\,^{\nu}}{(-\gamma)^{1/2}}+\rho \Sigma^{\mu \sigma} \nabla_{\mu}\big(\frac{\Sigma_{\sigma}\,\,^{\nu}}{(-\gamma)^{1/2}}\big) = 0.
 \end{equation}
which multiplication of Eq. (\ref{TS2}) by $\frac{\Sigma_{\nu \alpha}}{(-\gamma)^{1/2}}$ leads to $\nabla_{\mu}(\rho \Sigma ^{\mu \sigma})\Sigma _{\sigma}\,^{\nu}\Sigma_{\nu \alpha}/\gamma=0$. Contracting the previous identity with $\Sigma_{\alpha \nu}$ and using Eq. (\ref{Identity3}), we obtain $\nabla_{\mu}(\rho \Sigma ^{\mu \sigma})\Sigma_{\sigma}\,^{\nu}=0$, 
Finally by using a system of coordinates adapted to the parametrization of the surface, we get,
 \begin{equation}\label{T1}
 {\partial _\mu }(\sqrt { - g} \rho {\Sigma ^{\mu \sigma }}) = 0.
 \end{equation}
 By considering the following metric as an ansatz for a four-dimensional planar AdS black brane,
 \begin{equation}\label{metric}
 ds^{2} =-f(r)dt^{2} +\frac{dr^{2}}{f(r)} +r^2h_{ij}dx^idx^j,
 \end{equation}
  The equations of motion are given by,
  \begin{equation}\label{EoM}
  {G_{\mu \nu }} + \Lambda {g_{\mu \nu }} -{{m}^2}{\chi _{\mu \nu }} = {T^S _{\mu \nu }}
  \end{equation}
where $G_{\mu \nu }= R_{\mu \nu}-\frac{1}{2}g_{\mu \nu}R$  is the Einstein tensor, $T^S _{\mu \nu }$ is the energy-momentum tensor of matter that we consider as a cloud of string and  ${\chi _{\mu \nu }}$  is massive term,
  \begin{align}\label{chi}
  \mathcal{\chi}_{\mu \nu} =\frac{c_1}{2}\bigg(\mathcal{U}_1 g_{\mu \nu }-\mathcal{K}_{\mu \nu}\bigg)+\frac{c_2}{2}\bigg(\mathcal{U}_2 g_{\mu \nu }-2\mathcal{U}_1\mathcal{K}_{\mu \nu}+2\mathcal{K}^2_{\mu \nu}\bigg)+\frac{c_3}{2}\bigg(\mathcal{U}_3 g_{\mu \nu }-3\mathcal{U}_2 \mathcal{K}_{\mu \nu}\nonumber\\+6\mathcal{U}_1 \mathcal{K}^2_{\mu \nu}-6\mathcal{K}^3_{\mu \nu}\bigg)+\frac{c_4}{2}\bigg(\mathcal{U}_4g_{\mu \nu }-4\mathcal{U}_3 \mathcal{K}_{\mu \nu}+12\mathcal{U}_2 \mathcal{K}^2_{\mu \nu}-24\mathcal{U}_1 \mathcal{K}^3_{\mu \nu}+24\mathcal{K}^4_{\mu \nu}\bigg).
  \end{align}
Using this metric ansatz, $rr$  equation of motion Eq.(\ref{EoM}) reduces to,
\begin{equation}\label{eom}
rf'(r)+f(r)-k+\Lambda r^2- m^2\bigg(c_0c_1 r+c_0^2c_2\bigg)=r^2 T_r ^r
\end{equation}  
in which a prime denotes a differentiate with respect to the radial coordinate r. The density $\rho$ and the bivector $\Sigma$ are the function of $r$ for the spherically symmetric and static string cloud. The only non-vanishing component of the bivector $\Sigma$ is $\Sigma^{tr}=-\Sigma^{rt}$, since $\Sigma$ is a
bivector. So the surviving components of bivector are given by \cite{Herscovich:2010vr},
\begin{equation}
\Sigma^{\sigma \mu}=A(r)\bigg(\delta^{\sigma}_0 \delta^{\mu}_1-\delta^{\mu}_0 \delta^{\sigma}_1\bigg)
\end{equation} 
By plugging the non-zero components of $\Sigma^{\mu \nu}$ in Eq. (\ref{T}) we have
\begin{equation}\label{T^tt}
T^{tt}=-\frac{\rho}{\sqrt{-\gamma}}\Sigma^{tr}\Sigma_r\,^{t}
\end{equation} 
By inserting the value of $\gamma=\Sigma^{tr}\Sigma_{tr}=-(\Sigma^{tr})^2$ in Eq. (\ref{T^tt}), we get
\begin{equation}
 T_t ^t=T_r ^r=-\rho|A(r)|
\end{equation} 
 and from Eq. (\ref{T1}), we obtain ${\partial _r}(\frac{r^2}{l^2} T_t ^t) = 0$
which implies,
\begin{equation}
T^{\mu}_{ \nu}=-\frac{a}{r^2}diag[1,1,0,0]
\end{equation}
where $a$ is a positive constant.\\
By substituting the value of $T_r ^r$ in Eq. (\ref{eom}),
\begin{equation}
rf'(r)+f(r)-k+\Lambda r^2- m^2\bigg(c_0c_1 r+c_0^2c_2\bigg)=-a
\end{equation}
$f(r)$ is found as follows,
\begin{equation}\label{}
 f(r)=k-\frac{b}{r}-a-\frac{\Lambda}{3} r^2+m^2 \bigg(\frac{c_0c_1}{2}r+c_0^2c_2\bigg).
\end{equation}
 Event horizon is at $f(r_0)=0$ and we can find $b$ by applying this condition,
  \begin{equation}
b= r_0\Bigg[k-\frac{\Lambda}{3} r_0^2-a+m^2\bigg(\frac{c_0c_1}{2}r_0+c_0^2c_2\bigg)\Bigg]
 \equiv r_0\bigg(k-\frac{\Lambda}{3} r_0^2-a+\Delta\bigg)\\
\end{equation}
where $\Delta$ is,
 \begin{equation}
\Delta \equiv m^2\bigg(\frac{c_0c_1}{2}r_0+c_0^2c_2\bigg)
\end{equation}
In our case  $k$ is zero. By substituting $b$ in $f(r)$ we have,
 \begin{equation}\label{f}
f(r)=\frac{1}{r}\bigg[-a(r-r_0)-\frac{ \Lambda}{3}(r^3-r_0^3)+m^2c_0^2c_2(r-r_0)+m^2\frac{c_0c_1}{2}(r^2-r_0^2)\bigg].
\end{equation}
The entropy density can be found by applying Hawking-Bekenstein formula,
 \begin{equation}
s=\frac{A}{4G V_2}=\frac{4 \pi}{V_2} \int{d^2x \sqrt{-g}}=\frac{4 \pi}{V_2} \int{d^2x \sqrt{\chi}}=4 \pi \sqrt{\chi(r_0)}  =4\pi\frac{r_0^2}{l^2}
\end{equation}
where $V_{2}$ is the volume of the constant $t$ and $r$ hyper-surface with radius $r_{0}$ , $\chi(r_0)$ is the determinant of the spatial metric on the horizon and we used $\frac{1}{16\pi G} =1$ so $\frac{1}{4G} =4\pi$.
The temperature is
\begin{equation}
T=\frac{f'(r_0)}{4\pi}=\frac{1}{4\pi r_0}\bigg(\frac{3r_0^2}{l^2}-a +m^2l^2(r_0c_0c_1+c_0^2c_2)\bigg)
\end{equation}

\section{Holographic Aspects of the Solution}
\label{sec3}
 From the prescription of Green-Kubo formula (\ref{Kubo}) we can read the shear viscosity by 2-point function of energy-momentum tensor. According to AdS/CFT duality for calculating $[T{^x\,_y}(x),T{^x\,_y} (0)]$ we should perturb the bulk metric by $\delta g{^x\,_y}$.\\ 
We consider the metric and energy-momentum tensor as the following,  
\begin{align}\label{Background}
& ds^{2} =-g_{tt}(r)dt^{2} +g_{rr}(r)dr^{2}+g_{xx} (r) dx^i dx_i\\
& T_{\mu \nu}=diag \bigg(T_{tt}(r),T_{rr}(r),T_{xx}(r),T_{xx}(r)\bigg)
\end{align} 
they are homogeneous and isotropic in the field theory directions. By applying the perturbation $(\delta g){^x\,_y}=\phi(r)e^{-\imath\omega t}$ in the metric background (\ref{Background}) the decouple mode is,
\begin{equation}\label{mode}
\frac{1}{\sqrt{-g}}\partial_r\bigg(\sqrt{-g}g^{rr}\partial_r{\phi}\bigg)+[g^{tt}\omega^2-m(r)^2]\phi=0
\end{equation}
\begin{equation}
m(r)^2=g^{xx}T_{xx}-\frac{\delta T_{xy}}{\delta g_{xy}}\nonumber
\end{equation}
Shear viscosity is calculated by Eq.(\ref{Kubo}),\\
\begin{equation}
\eta=\mathop{\lim }\limits_{\omega \, \to \, 0} \frac{1}{\omega } \Im G^{R} _{T^{xy}T^{xy}} (\omega ,k=0)=\frac{\sqrt{\chi(r_0)}}{16\pi G_N}\phi_0(r_0)^2=\frac{s}{4\pi} \phi_0(r_0)^2
\end{equation}
 Then, we will have,
\begin{equation}\label{formula}
\frac{\eta}{s}=\frac{1}{4\pi} \phi_0(r_0)^2
\end{equation}
where $\phi_0$ is the solution of mode equation  (\ref{mode}) at zero frequency ($\omega=0$).\\ $\phi$ has these 2 conditions: (i) $\phi$ is regular at horizon $r=r_0$ and (ii) goes like to $\phi=1$ near the boundary as $r \to \infty$.\\
For calculation of shear viscosity from Green-Kubo formula we should perturb the metric (\ref{metric}) according to above procedure,
\begin{equation}
ds^2=-\frac{f_1(r)}{l^2}dt^2 + \frac{l^2}{f_1(r)}dr^2 + \frac{r^2}{l^2}(dx^2 + dy^2 + 2\phi(r)dxdy)
\end{equation}
\begin{align}
f_1(r)&=\frac{l^2}{r}\bigg[-a(r-r_0)-\frac{\Lambda}{3}(r^3-r_0^3)+m^2 c_0^2c_2(r-r_0)\nonumber\\&+m^2\frac{c_0c_1}{2}(r^2-r_0^2)\bigg]=l^2 f(r)
\end{align}
By choosing the metric perturbation as $\delta g_{xy}=\frac{r^2}{l^2}\phi(r)e^{i\omega t}$ and inserting it into the action Eq.(\ref{action}) and keeping up to $\phi^2$ \cite{Kovtun2012,Son2007,Policastro2001,Kovtun2004,Policastro2002,Kovtun2003}, we get:
\begin{equation}\label{perturbed}
S_2=\frac{-1}{2} \int d^4x \Big(K_1 \phi'^2 -K_2 \phi^2\Big)
\end{equation}
we demand $\omega=0$, where
\begin{align}
K_1&=\frac{r^2 f_1(r)}{l^4}=\frac{r^2 f(r)}{l^2} \nonumber\\
&=\frac{r(r-r_0)}{2l^4}\Big(-2a+2 c_0^2 c_2 l^2 m^2 + c_0 c_1 l^2 m^2 (r+r_0)\Big)+\frac{r(r^3-r_0^3)}{l^2} \nonumber\\
K_2&=\frac{c_0c_1m^2 r}{2l^2}
\end{align}
then the EoM is,
\begin{equation}\label{EoM2}
(K_1 \phi')' + K_2 \phi=0
\end{equation}
We try to solve the mode equation Eq.(\ref{EoM2}) perturbatively in $m^2$ and  $a$. So firstly consider $m=a=0$. Then the EoM will be given as follows,
\begin{equation}\label{EoM-zeromass1}
r(r^3-r_0^3)\phi^{''} + (4r^3-r_0^3) \phi' =0
\end{equation}
The solution is
\begin{equation}
\phi(r)=C_2+C_1 (-\log r+\log(r-r_0)).
\end{equation}
Then applying the boundary conditions (regularity at horizon and $\phi=1$ at the boundary) gives $C_2=0$ and $C_1=1$ which means that $\phi(r)=1$ is a constant solution. In this case, according to Eq. (\ref{formula}),
\begin{equation}\label{ValuM,a=0}
\frac{\eta}{s}= \frac{1}{4\pi}\phi(r_0)^2 = \frac{1}{4\pi}
\end{equation}
 Now consider $m^2$  and $a$ to be a small parameter and try to solve Eq.(\ref{EoM2}). By Putting $\phi=\phi_0+m^2\phi_1(r)+a \phi_2(r)$ where $\phi_0=1$ and expanding EoM in terms of powers of $m^2$ and $a$, we will find,
  \begin{equation}
  m^2\Big( c_0 c_1 L^2 r+2 r (r^3 - r_0^3) \phi''_1(r)+2(4 r^3- r_0^3) \phi'_1(r)  \Big)=0
  \end{equation} 
  \begin{equation}
a\bigg(r(r^3-r_0^3)\phi_2^{''} + (4r^3-r_0^3) \phi_2'\bigg)  =0
  \end{equation}  
 Thus we find the solutions,
 \begin{align}\label{Phi1}
 \phi_1(r)&=C_2 -\frac{m^2}{24 r_0^3}\Big( 2\sqrt{3}c_0 c_1 l^2 r_0^2 ArcTan(\frac{2r+r_0}{\sqrt{3}r_0}) +\nonumber\\&+24 C_1 \log r + 2c_0 c_1 l^2 r_0^2 \log(r_0-r)\nonumber\\& - c_0c_1l^2 r_0^2 \log(r^2+r r_0 +r_0^2) - 8C_1 \log(r-r_0) -8C_1\log(r^2+r r_0 +r_0^2) \Big)
 \end{align}
 \begin{equation}\label{Phi2}
 \phi_2(r)=C_3+\frac{C_4}{3r_0^3}\bigg(\log(r-r_0)+\log(r^2+r r_0 +r_0^2)-3\log r\bigg)
 \end{equation}
 $\phi(r)$ is as follows,  
 \begin{align}\label{Phi}
 \phi(r)&=\Phi_0- \frac{m^2}{24 r_0^3}\Big( 2\sqrt{3}c_0 c_1 l^2 r_0^2 ArcTan(\frac{2r+r_0}{\sqrt{3}r_0}) +\nonumber\\&+24 C_1 \log r + 2c_0 c_1 l^2 r_0^2 \log(r_0-r)\nonumber\\& - c_0c_1l^2 r_0^2 \log(r^2+r r_0 +r_0^2) - 8C_1 \log(r-r_0) -8C_1\log(r^2+r r_0 +r_0^2) \Big)\nonumber\\&+\frac{aC_4}{3r_0^3}\bigg(\log(r-r_0)+\log(r^2+r r_0 +r_0^2)-3\log r\bigg)
 \end{align}
 where $\Phi_0=\phi_0+a C_3+m^2 C_2$.\\
 By applying regularity on horizon condition, we will get rid of $\log(r-r_0)$ if we choose $C_1=\frac{1}{4}r_0^2l^2 c_1 c_0 -\frac{aC_4}{m^2}$.\\
The second boundary condition is at $\phi(r=\infty)=1$, which gives,
\begin{equation}
\Phi_0= 1+\frac{m^2 l^2 \pi c_0c_1}{24 r_0} ( \sqrt{3} + 2i)
\end{equation}
Thus we have found $\phi(r)$ from horizon to boundary. The solution is as follows,
\begin{align}
\phi(r)=1+\frac{m^2 l^2 c_0c_1}{24 r_0} \bigg(\sqrt{3}\pi -2 \sqrt{3}ArcTan(\frac{2r+r_0}{\sqrt{3}r_0})-6\log r +3 \log (r^2+rr_0+r_0^2) \bigg)
\end{align}
So we have solved the mode equation Eq.(\ref{EoM2}) up to the first order in $a$ and $m^2$. In order to calculate $\frac{\eta}{s}$ we should find the value of $\phi(r)$ at $r=r_0$ 
 \begin{align}
 \frac{\eta}{s}&= \frac{1}{4\pi}\phi(r_0)^2 = \frac{1}{4\pi} \bigg(1+\frac{m^2 c_0 c_1 l^2}{36 r_0}(\sqrt{3}\pi +9 \log3) \bigg)
 \end{align}\\\\
 KSS bound is violated by applying regularity of the solution of mode equation at horizon and $\phi(r)=1$ on the boundary of AdS.
 
The violation of KSS in higher derivative gravity is not arbitrary. Causality sets a different and lower bound but $\frac{\eta}{s}$ cannot vanishes to 0 continuously. In massive gravity the case is different, at low temperature the $\frac{\eta}{s}$ ratio goes to zero in a power law fashion\cite{Hartnoll:2016tri}. From a technical point of view there is another important difference. Massive gravity theories violate the KSS bound due to the appearance of a mass term in the equation for the shear component of the graviton (\ref{mode}). This is not true in the higher derivative gravity where the reason of the violation is very different.\\
Momentum dissipation not always implies the violation of the KSS bound \cite{Alberte:2016xja}. There are fluids theory, which are introduced in \cite{Alberte:2015isw} and studied in detail in \cite{Baggioli:2019abx}, which dissipate momentum (like your dRGT theory) but they do not violate the KSS bound.
Similar results are also presented  in \cite{Baggioli:2018bfa} from both analytical and numerical aspects. 
 \section{Conclusion}
\noindent We study the aspects of field theory sector of massive gravity with a cloud of strings. $\frac{\eta}{s}$ is calculated  by applying the Dirichlet boundary and regularity on the horizon conditions\cite{Hartnoll:2016tri}. This value is an important quantity in fluid-gravity duality and proportional to the inverse squared of the coupling of the field theory sector, so it means that the field theory side of massive gravity with a cloud of strings is the same as massive gravity theory. There is a conjecture that states $\frac{\eta }{s}=\frac{1}{4\pi}$ for Einstein-Hilbert gravity, known as KSS bound \cite{Ref22} which it is violated for higher derivative gravity\cite{Ref22,Ref23,Ref24,Sadeghi:2015vaa,Parvizi:2017boc,Sadeghi:2018vrf}. Our result shows massive gravity with a cloud of string behaves like pure AdS massive gravity\cite{Sadeghi:2018ylh}. Our outcome also shows that a cloud of strings acts like a charge comparison to \cite{Sadeghi:2018ylh,Wu:2018zoq} results.\\\\
\noindent {\large {\bf Acknowledgment} }  The authors would like to thank Shahrokh Parvizi and Matteo Baggioli for valuable suggestions and comments.


\begin{thebibliography}{}

\bibitem{Dvali2000}
G. Dvali, G. Gabadadze and M. Porrati, Phys. Lett. B 484, 112 (2000).

\bibitem{Dvali2000-}
G. Dvali, G. Gabadadze and M. Porrati, Phys. Lett. B 485, 208 (2000).


\bibitem{deRham:2010kj} 
C.~de Rham, G.~Gabadadze and A.~J.~Tolley,
Phys.\ Rev.\ Lett.\  {\bf 106}, 231101 (2011)
[arXiv:1011.1232 [hep-th]].

\bibitem{Hassan:2011ea} 
S.~F.~Hassan and R.~A.~Rosen,
``Confirmation of the Secondary Constraint and Absence of Ghost in Massive Gravity and Bimetric Gravity,''
JHEP {\bf 1204}, 123 (2012)
[arXiv:1111.2070 [hep-th]].


\bibitem{Gergely2006} 
L.~A.~Gergely,
   ``Brane-world cosmology with black strings,''  Phys.\ Rev.\ D {\bf 74}, 024002 (2006)  [hep-th/0603244].

\bibitem{Brans1961}
C.~Brans and R.~H.~Dicke,
``Mach's principle and a relativistic theory of gravitation,''  Phys.\ Rev.\  {\bf 124}, 925 (1961).

\bibitem{Akbar}
M.~Akbar and R.~G.~Cai,
``Thermodynamic Behavior of Field Equations for f(R) Gravity,''  Phys.\ Lett.\ B {\bf 648}, 243 (2007)  [gr-qc/0612089].

\bibitem{Lovelock:1971}
D.~Lovelock,
``The Einstein tensor and its generalizations,''  J.\ Math.\ Phys.\  {\bf 12}, 498 (1971).






\bibitem{Vasiliev96}
M. A. Vasiliev, Int. J. Mod. Phys. D 5, 763 (1996).


\bibitem{Fierz:1939ix} 
M.~Fierz and W.~Pauli,
``On relativistic wave equations for particles of arbitrary spin in an electromagnetic field,''
Proc.\ Roy.\ Soc.\ Lond.\ A {\bf 173}, 211 (1939).

\bibitem{Boulware72}
D. G. Boulware and S. Deser, Phys. Rev. D 6, 3368 (1972).

\bibitem{Maldacena97}
J. M. Maldacena, ``The Large N limit of superconformal field theories and supergravity,'' Int.\ J.\ Theor.\ Phys.\  {\bf 38} (1999) 1113 [Adv.\ Theor.\ Math.\ Phys.\  {\bf 2} (1998) 231] [hep-th/9711200].

\bibitem{Aharony99}
O. Aharony, S.~S.~Gubser, J.~M.~Maldacena, H.~Ooguri and Y.~Oz,
``Large N field theories, string theory and gravity,''
Phys.\ Rept.\  {\bf 323}, 183 (2000)
[hep-th/9905111].

\bibitem{Casalderrey11}
J.~Casalderrey-Solana, H.~Liu, D.~Mateos, K.~Rajagopal and U.~A.~Wiedemann,
``Gauge/String Duality, Hot QCD and Heavy Ion Collisions,''  arXiv:1101.0618 [hep-th].

\bibitem{Mateos07}
D.~Mateos,
``String Theory and Quantum Chromodynamics,''
Class.\ Quant.\ Grav.\  {\bf 24}, S713 (2007)
[arXiv:0709.1523 [hep-th]].

\bibitem{Bhattacharyya07}
S.~Bhattacharyya, V.~E.~Hubeny, S.~Minwalla and M.~Rangamani,
``Nonlinear Fluid Dynamics from Gravity,''
JHEP {\bf 0802}, 045 (2008)
[arXiv:0712.2456 [hep-th]].



\bibitem{Bhattacharya2011}
J.~Bhattacharya, S.~Bhattacharyya, S.~Minwalla and A.~Yarom,
``A Theory of first order dissipative superfluid dynamics,''
JHEP {\bf 1405}, 147 (2014)
[arXiv:1105.3733 [hep-th]].


\bibitem{Kovtun2012}
P. Kovtun,``Lectures on hydrodynamic fluctuations in relativistic theories,''J.\ Phys.\ A {\bf 45} (2012) 473001[arXiv:1205.5040 [hep-th]].



\bibitem{Son2007}
D.~T.~Son and A.~O.~Starinets,
``Viscosity, Black Holes, and Quantum Field Theory,''
Ann.\ Rev.\ Nucl.\ Part.\ Sci.\  {\bf 57}, 95 (2007)
[arXiv:0704.0240 [hep-th]].

\bibitem{Policastro2001}
G.~Policastro, D.~T.~Son and A.~O.~Starinets,
``The Shear viscosity of strongly coupled $ \mathcal{N}=4 $ supersymmetric Yang-Mills plasma,''
Phys.\ Rev.\ Lett.\  {\bf 87}, 081601 (2001)
[hep-th/0104066].

\bibitem{Kovtun2004}
P.~Kovtun, D.~T.~Son and A.~O.~Starinets,
``Viscosity in strongly interacting quantum field theories from black hole physics,''
Phys.\ Rev.\ Lett.\  {\bf 94}, 111601 (2005)
[hep-th/0405231].

\bibitem{Policastro2002}
G.~Policastro, D.~T.~Son and A.~O.~Starinets,``From AdS/CFT correspondence to hydrodynamics,''
JHEP {\bf 0209}, 043 (2002)
[hep-th/0205052]. 

\bibitem{Sadeghi:2019trg} 
M.~Sadeghi,
``Transverse Shear Viscosity to Entropy Density for the General Anisotropic Black Brane in Horava-Lifshitz Gravity,''
doi:10.1007/s12648-019-01523-6
arXiv:1905.02932 [hep-th].

\bibitem{Nojiri:2010ny} 
S.~Nojiri and S.~D.~Odintsov,
``Non-singular modified gravity unifying inflation with late-time acceleration and universality of viscous ratio bound in F(R) theory,''
Prog.\ Theor.\ Phys.\ Suppl.\  {\bf 190}, 155 (2011)
doi:10.1143/PTPS.190.155
[arXiv:1008.4275 [hep-th]].


\bibitem{Kovtun2003}
P.~Kovtun, D.~T.~Son and A.~O.~Starinets,
``Holography and hydrodynamics: Diffusion on stretched horizons,''
JHEP {\bf 0310}, 064 (2003)
[hep-th/0309213].

\bibitem{Letelier:1979ej} 
  P.~S.~Letelier,
  ``Clouds Of Strings In General Relativity,''
  Phys.\ Rev.\ D {\bf 20}, 1294 (1979).
  doi:10.1103/PhysRevD.20.1294.


\bibitem{Richarte:2007bx} 
  M.~G.~Richarte and C.~Simeone,
  ``Traversable wormholes in a string cloud,''
  Int.\ J.\ Mod.\ Phys.\ D {\bf 17}, 1179 (2008)
  doi:10.1142/S0218271808012759
  [arXiv:0711.2297 [gr-qc]].

\bibitem{Yadav:2009zza} 
  A.~K.~Yadav, V.~K.~Yadav and L.~Yadav,
  ``Cylindrically symmetric inhomogeneous universes with a cloud of strings,''
  Int.\ J.\ Theor.\ Phys.\  {\bf 48}, 568 (2009)
  doi:10.1007/s10773-008-9832-9
  [arXiv:1112.4114 [gr-qc]].

\bibitem{Ganguly:2014cqa} 
  A.~Ganguly, S.~G.~Ghosh and S.~D.~Maharaj,
 ``Accretion onto a black hole in a string cloud background,''
  Phys.\ Rev.\ D {\bf 90}, no. 6, 064037 (2014)
  doi:10.1103/PhysRevD.90.064037
  [arXiv:1409.7872 [gr-qc]].

\bibitem{Bronnikov:2016dhz} 
  K.~A.~Bronnikov, S.~W.~Kim and M.~V.~Skvortsova,
``The Birkhohff theorem and string clouds,''
  Class.\ Quant.\ Grav.\  {\bf 33}, no. 19, 195006 (2016)
  doi:10.1088/0264-9381/33/19/195006
  [arXiv:1604.04905 [gr-qc]].

\bibitem{Barbosa:2016lse} 
  D.~Barbosa and V.~B.~Bezerra,
``On the rotating Letelier spacetime,''
  Gen.\ Rel.\ Grav.\  {\bf 48}, no. 11, 149 (2016).
  doi:10.1007/s10714-016-2143-1.

\bibitem{Herscovich:2010vr} 
  E.~Herscovich and M.~G.~Richarte,
  ``Black holes in Einstein-Gauss-Bonnet gravity with a string cloud background,''
  Phys.\ Lett.\ B {\bf 689}, 192 (2010)
  doi:10.1016/j.physletb.2010.04.065
  [arXiv:1004.3754 [hep-th]].

\bibitem{Ghosh:2014pga} 
  S.~G.~Ghosh, U.~Papnoi and S.~D.~Maharaj,
  ``Cloud of strings in third order Lovelock gravity,''
  Phys.\ Rev.\ D {\bf 90}, no. 4, 044068 (2014)
  doi:10.1103/PhysRevD.90.044068
  [arXiv:1408.4611 [gr-qc]].

\bibitem{Lee:2014dha} 
  T.~H.~Lee, D.~Baboolal and S.~G.~Ghosh,
 ``Lovelock black holes in a string cloud background,''
  Eur.\ Phys.\ J.\ C {\bf 75}, no. 7, 297 (2015)
  doi:10.1140/epjc/s10052-015-3515-5
  [arXiv:1409.2615 [gr-qc]].

\bibitem{Mazharimousavi:2015sfo} 
  S.~H.~Mazharimousavi and M.~Halilsoy,
  ``Cloud of strings as source in $2+1$ -dimensional $f\left( R\right) =R^{n}$ gravity,''
  Eur.\ Phys.\ J.\ C {\bf 76}, no. 2, 95 (2016)
  doi:10.1140/epjc/s10052-016-3954-7
  [arXiv:1511.00603 [gr-qc]].

\bibitem{Graca:2016cbd} 
  J.~P.~Morais Graça, G.~I.~Salako and V.~B.~Bezerra,
  ``Quasinormal modes of a black hole with a cloud of strings in Einstein-Gauss-Bonnet gravity,''
  Int.\ J.\ Mod.\ Phys.\ D {\bf 26}, no. 10, 1750113 (2017)
  doi:10.1142/S0218271817501139
  [arXiv:1604.04734 [gr-qc]].




      
        


\bibitem{Herscovich:2010vr} 
E.~Herscovich and M.~G.~Richarte,
``Black holes in Einstein-Gauss-Bonnet gravity with a string cloud background,'' 
 Phys.\ Lett.\ B {\bf 689}, 192 (2010)  doi:10.1016/j.physletb.2010.04.065  [arXiv:1004.3754 [hep-th]].

\bibitem{Cai:2014znn} 
R.~G.~Cai, Y.~P.~Hu, Q.~Y.~Pan and Y.~L.~Zhang,
``Thermodynamics of Black Holes in Massive Gravity,''
Phys.\ Rev.\ D {\bf 91}, no. 2, 024032 (2015)
[arXiv:1409.2369 [hep-th]].

 

\bibitem{Hartnoll:2016tri} 
S.~A.~Hartnoll, D.~M.~Ramirez and J.~E.~Santos,
``Entropy production, viscosity bounds and bumpy black holes,''
JHEP {\bf 1603}, 170 (2016)
doi:10.1007/JHEP03(2016)170
[arXiv:1601.02757 [hep-th]].



\bibitem{Alberte:2016xja} 
  L.~Alberte, M.~Baggioli and O.~Pujolas,
  ``Viscosity bound violation in holographic solids and the viscoelastic response,''
  JHEP {\bf 1607}, 074 (2016)
  doi:10.1007/JHEP07(2016)074
  [arXiv:1601.03384 [hep-th]].

\bibitem{Alberte:2015isw} 
  L.~Alberte, M.~Baggioli, A.~Khmelnitsky and O.~Pujolas,
  ``Solid Holography and Massive Gravity,''
  JHEP {\bf 1602}, 114 (2016)
  doi:10.1007/JHEP02(2016)114
  [arXiv:1510.09089 [hep-th]].



\bibitem{Baggioli:2019abx} 
  M.~Baggioli and S.~Grieninger,
  ``Zoology of Solid and Fluid Holography : Goldstone Modes and Phase Relaxation,''
  arXiv:1905.09488 [hep-th].

\bibitem{Baggioli:2018bfa} 
  M.~Baggioli and A.~Buchel,
  ``Holographic Viscoelastic Hydrodynamics,''
  JHEP {\bf 1903}, 146 (2019)
  doi:10.1007/JHEP03(2019)146
  [arXiv:1805.06756 [hep-th]].

\bibitem{Ref22}
M.~Brigante, H.~Liu, R.~C.~Myers, S.~Shenker and S.~Yaida,
  ``Viscosity Bound Violation in Higher Derivative Gravity,''
  Phys.\ Rev.\ D {\bf 77}, 126006 (2008)
  [arXiv:0712.0805 [hep-th]]. 

 
  
\bibitem{Ref23}
M.~Brigante, H.~Liu, R.~C.~Myers, S.~Shenker and S.~Yaida,
   ``The Viscosity Bound and Causality Violation,''
   Phys.\ Rev.\ Lett.\  {\bf 100}, 191601 (2008)
   [arXiv:0802.3318 [hep-th]].

 \bibitem{Ref24}
  I.~P.~Neupane and N.~Dadhich,
     ``Entropy Bound and Causality Violation in Higher Curvature Gravity,''
     Class.\ Quant.\ Grav.\  {\bf 26}, 015013 (2009)
     [arXiv:0808.1919 [hep-th]].

 

\bibitem{Sadeghi:2015vaa} 
  M.~Sadeghi and S.~Parvizi,
  ``Hydrodynamics of a black brane in Gauss-Bonnet massive gravity,''
  Class.\ Quant.\ Grav.\  {\bf 33}, no. 3, 035005 (2016)
  [arXiv:1507.07183 [hep-th]].


\bibitem{Parvizi:2017boc} 
S.~Parvizi and M.~Sadeghi,
``Holographic Aspects of a Higher Curvature Massive Gravity,''
Eur.\ Phys.\ J.\ C {\bf 79}, no. 2, 113 (2019)
doi:10.1140/epjc/s10052-019-6631-9
[arXiv:1704.00441 [hep-th]].

\bibitem{Sadeghi:2018vrf} 
M.~Sadeghi,
``Black Brane Solution in Rastall AdS Massive Gravity and Viscosity Bound,''
Mod.\ Phys.\ Lett.\ A {\bf 33}, no. 37, 1850220 (2018)
doi:10.1142/S0217732318502206
[arXiv:1809.08698 [hep-th]].

\bibitem{Sadeghi:2018ylh} 
M.~Sadeghi,
``Einstein-Yang-Mills AdS black brane solution in massive gravity and viscosity bound,''
Eur.\ Phys.\ J.\ C {\bf 78}, no. 10, 875 (2018)
doi:10.1140/epjc/s10052-018-6360-5
[arXiv:1810.09242 [hep-th]].

\bibitem{Wu:2018zoq} 
B.~Wu and D.~C.~Zou,
``Viscosity/entropy ratio in the context of dRGT massive gravity,''
EPL {\bf 124}, no. 2, 20002 (2018).
doi:10.1209/0295-5075/124/20002.

  
    
      








\end{thebibliography}
\end{document}